\documentclass[12pt]{iopart}

\usepackage{graphicx}
\usepackage{epstopdf}
\newcommand{\Npart}{$\langle N_{part}\rangle$}

\newcommand{\sNN}[1]{$\sqrt{s_{NN}} = #1$ GeV}
\newcommand{\Lam}{$\Lambda $ }
\newcommand{\ALam}{$\bar{\Lambda} $ }
\newcommand{\Kaon}{$K^{0}_{S} $ }

\begin{document}

\title[Overview of Strangeness Production at the STAR Experiment]{Overview of Strangeness Production at the STAR Experiment}

\author{Anthony R. Timmins for the STAR Collaboration}

\address{Department of Physics and Astronomy, Wayne State University, 666 W. Hancock, Detroit, MI 48201, USA}
\ead{tone421@rcf.rhic.bnl.gov}
\begin{abstract}

We present an overview of recent STAR results on strangeness production in p+p and heavy-ion collisions at RHIC. In both Cu+Cu and Au+Au collisions we show the centrality dependencies of bulk yield and mid-$p_{T}$ spectrum measurements with new comparisons to theory. The latest $v_{2}$ results for strange particles are presented and prospects for strangeness production in the low energy scan program will be outlined. Finally, we report new measurements of strangeness fragmentation functions for jets in p+p collisions. 

\end{abstract}


\section{Introduction}

Measurements of strangeness production in heavy-ion collisions were originally conceived to be the smoking gun of Quark Gluon Plasma (QGP) formation \cite{RafalMull1, RafalMull2}. It was argued that due to a drop in the strange quark's dynamical mass and increased production cross section, strangeness in the QGP would saturate on small time scales relative to a hadronic gas. Twenty nine years since then, the production of strangeness continues to test our understanding of QCD in both p+p and heavy-ion collisions. In these proceedings, we review recent measurements of strange particle production at the STAR experiment. We first explore integrated yields in Cu+Cu and Au+Au collisions where production is expected to be dominated by low $q^2$ processes at RHIC energies, then proceed to mid-$p_T$ spectra where fragmentation and recombination may compete. We then turn to strange particle $v_2$ and tests for non-equilibrium behaviour with respect to elliptic flow, then show test results for RHIC's low energy program which will search for a possible critical point on the QCD phase diagram. Finally, we show new results on strange particle fragmentation functions in p+p collisions.

\section{Bulk Strangeness Production}

\begin{figure}[t]
\begin{center}
\includegraphics[width = 1\textwidth]{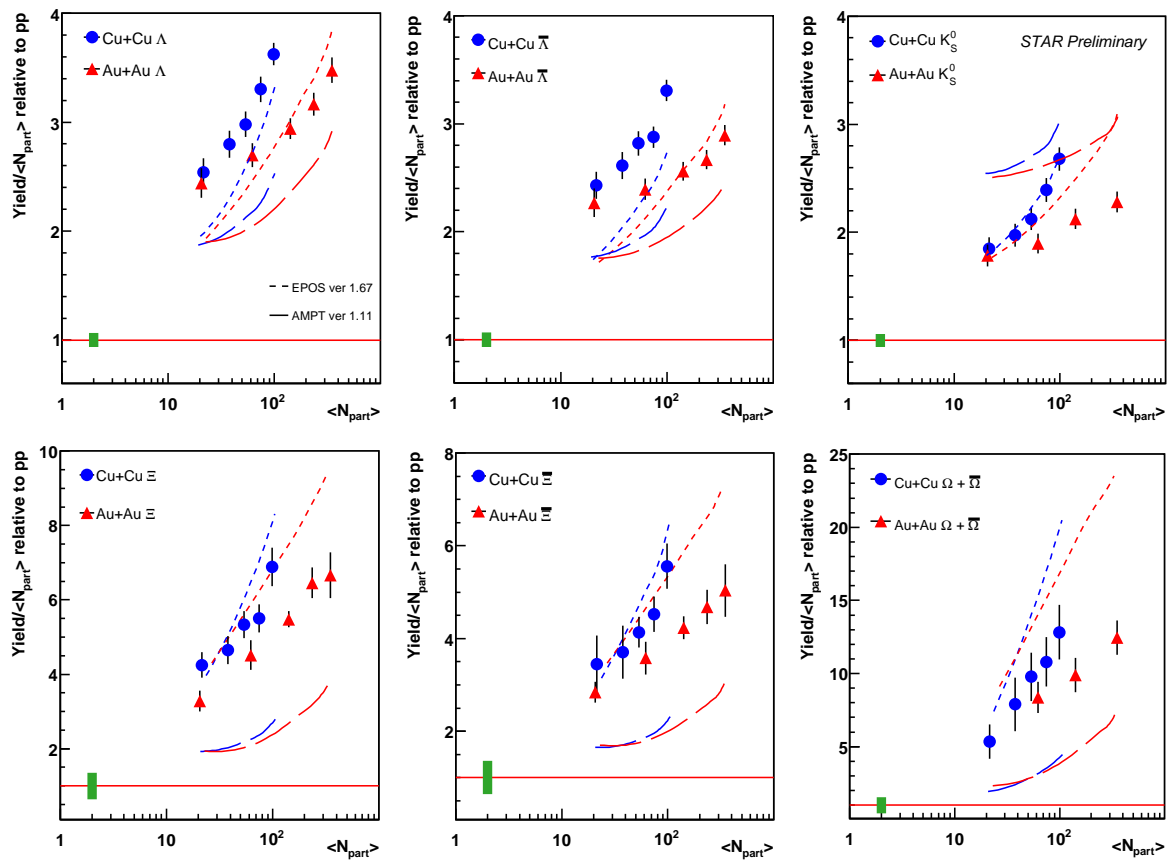}
\end{center}
\caption{Mid-rapidity per participant yields ($dN/dy$ per \Npart) of singly and multi strange particles for Cu+Cu and Au+Au collisions with \sNN{200} divided by the respective p+p values. The $\Lambda$ and $\bar{\Lambda}$ yields have been feed down subtracted in all cases. The green bars show the normalisation uncertainties, and the uncertainties for the heavy-ion points are the combined statistical and systematic. The dotted and dashed lines show EPOS and AMPT predictions respectively where the p+p reference corresponds to the experimentally measured yields \cite{MyHotQuark}.} 
\label{fig:1} 
 \end{figure}

Figure \ref{fig:1} shows the \emph{enhancement factor} for strange particle yields in Cu+Cu and Au+Au  \sNN{200} collisions. Assuming a thermally equilibrated QGP hadronises into a maximum entropy state, a test for strange quark saturation in the initial stages is provided by comparing the final hadron state yields to thermal model predictions from the Canonical formalism \cite{CanoncialSuppression}. For all strange particles it is clear that for a given system, rises in the enhancement factor with $\langle N_{part}\rangle$ are observed which is predicted by the Canonical formalism. However, the enhancement factor appears to rise more rapidly in Cu+Cu collisions, and this is inconsistent with the Canonical formalism which predicts the rise should just be controlled by $N_{part}$ independent of the collision system. That relies on a constant baryo-chemical potential and chemical freeze out temperature for both systems which has been shown \cite{AnetaPri, JunsPres}. 
\begin{figure}[h]
\begin{center}
\includegraphics[width = 0.45\textwidth]{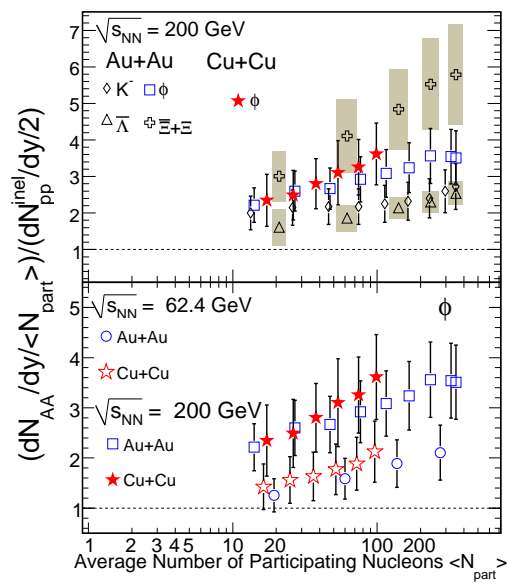}
\end{center}
\caption{The $\phi$ enhancement factor in heavy-ion collisions. The uncertainties on the data points are the combined statistical and systematic \cite{PHiCuAu}. }
\label{fig:2} 
\end{figure}
We also make comparisons to predictions from the AMPT \cite{AMPT} and EPOS \cite{EPOS} models which have been shown to describe the bulk features of hadron production well for Au+Au \sNN{200} collisions. The AMPT model is based on HIJING, and thus describes particle production in heavy-ion collisions via intra-nucleon string excitation and breaking (soft), and mini-jet fragmentation (hard) where the excited nucleons fragment independently. The EPOS model describes particle production with core and corona contributions. The core occupies the high density region in the collision zone and aims to mimic various QGP behaviour. At a particular freeze-out density, the hadronisation is treated via statistical emission where strangeness production is over saturated ($\gamma_S = 1.3$). Corona production occurs in the low density region and is a superposition of p+p collisions. Both the AMPT and EPOS models reproduce three key qualitative aspects of the data: rises in yields per $\langle N_{part}\rangle$ for a given system, a faster rise in yields per $\langle N_{part}\rangle$ for Cu+Cu collisions, and a merging in yields per $\langle N_{part}\rangle$ for peripheral Cu+Cu  and very peripheral Au+Au collisions. Quantitatively, EPOS is nearly always closer to the measured data than AMPT. 
Figure \ref{fig:2} shows the enhancement factor for the $\phi$ particle in context of the $K^{-}$, \ALam and $\Xi$ enhancements. For both Cu+Cu and Au+Au collisions at energies of $\sqrt{s_{NN}} =62$ and $200$ GeV, an above unity enhancement factor is observed which is also inconsistent with predictions from the Canonical formalism \cite{phiEnSQM, phiCoreCorona1}, and this sits between the single and multi-strange values for Au+Au \sNN{200} collisions.  An alternative core-corona approach has been shown to describe the $\phi$ enhancement factors in Au+Au \sNN{200} collisions quite well \cite{phiCoreCorona1}, as has a more recent implementation of the EPOS model \cite{phiCoreCorona2}.

\section{Thermal Model Comparisons}

\begin{figure}[h]
\begin{tabular}{c c}
\includegraphics[width = 0.48\textwidth]{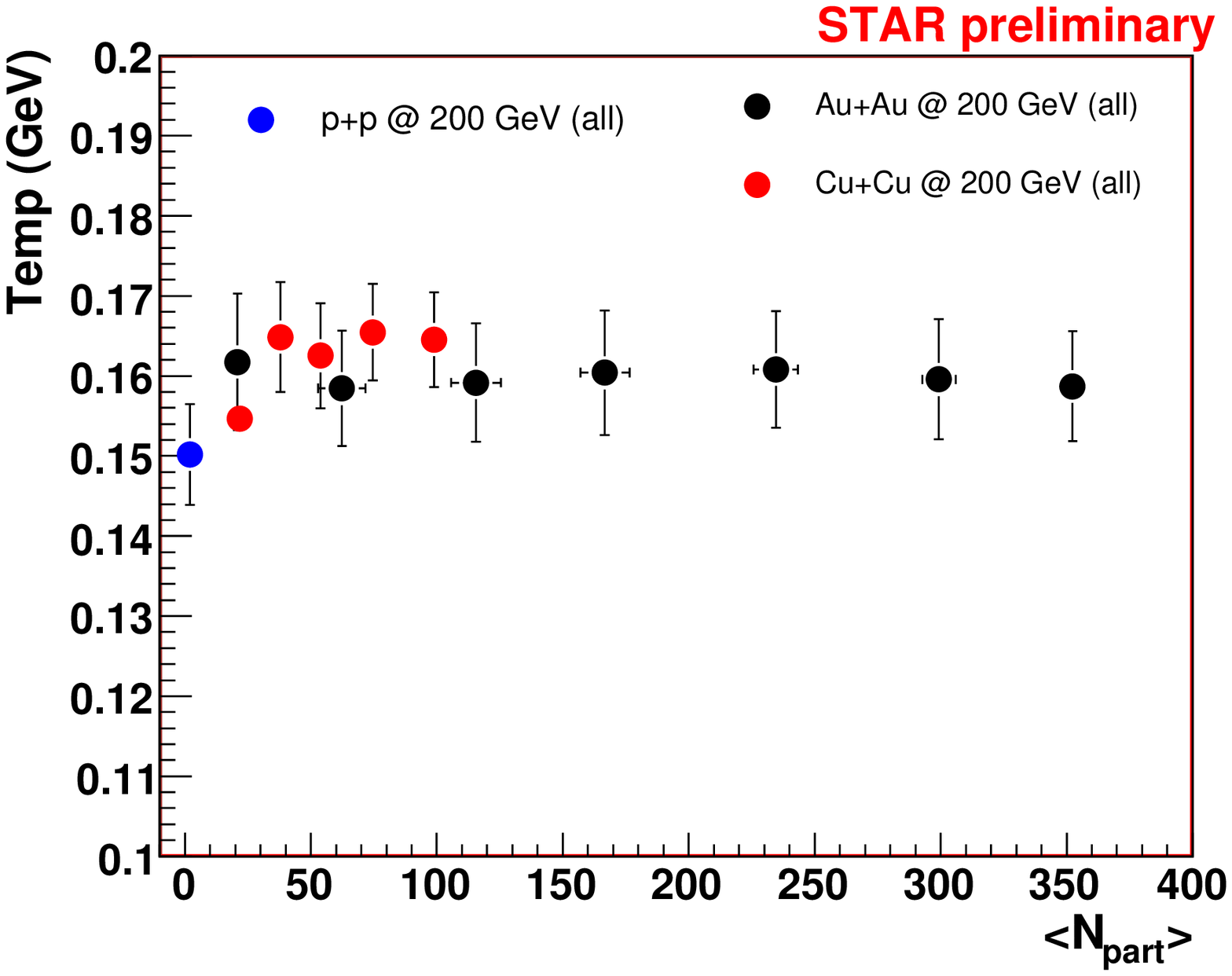}
&
\includegraphics[width = 0.48\textwidth]{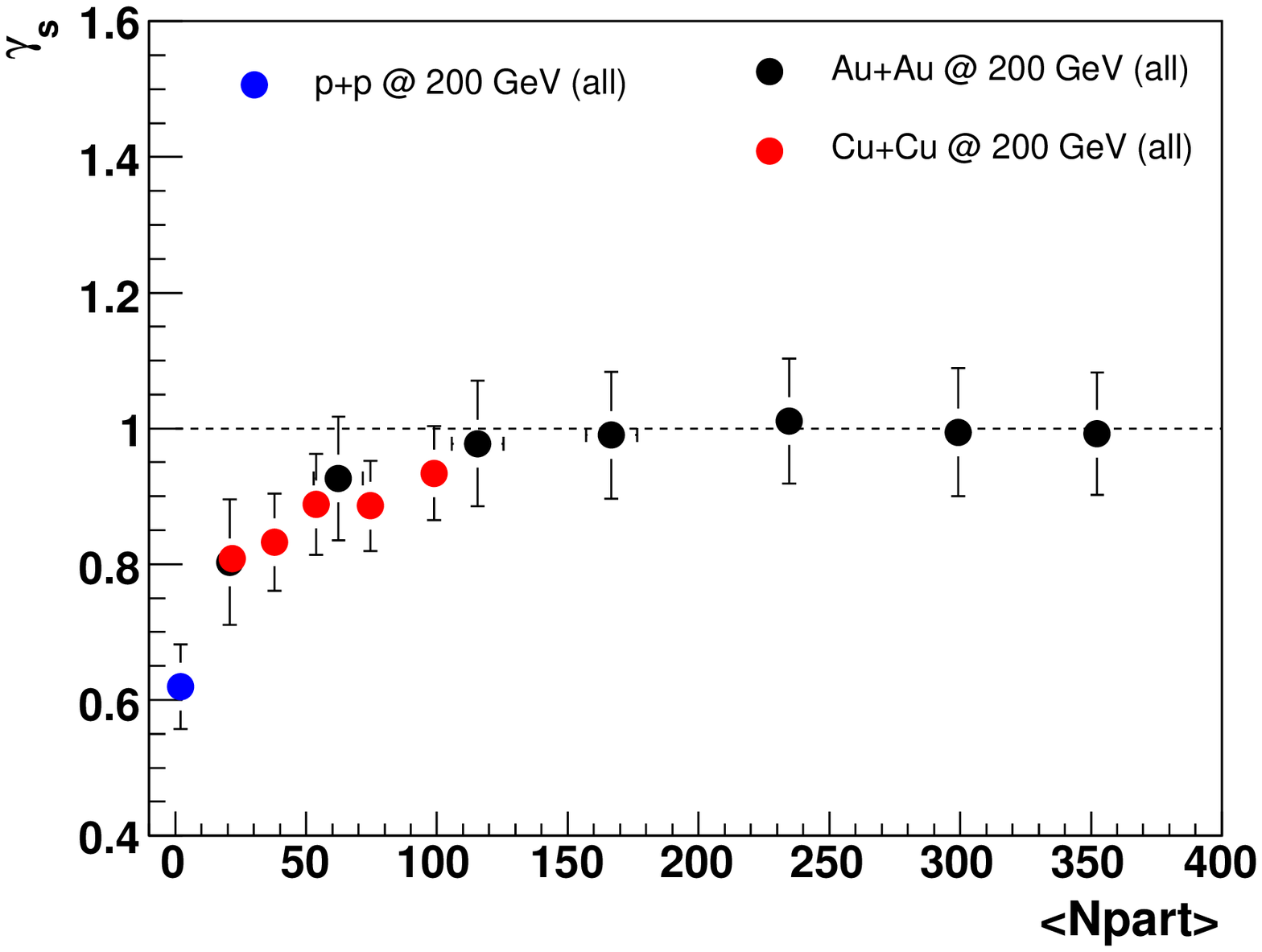}
\end{tabular}
\caption{Chemical freeze-out temperature ($T_{ch}$) and the strangeness saturation factor ($\gamma_{S}$) as a function of the number of participants for all particles measured by STAR (\Npart)\cite{JunsPres}. }
\label{fig:3} 
\end{figure}

Figure \ref{fig:3} shows thermal parameters extracted from particle ratios via the THERMUS model \cite{THERMUS} for p+p, Cu+Cu, and Au+Au \sNN{200} collisions. In the left panel we observe that the chemical freeze-out temperature ($T_{ch}$)  is independent of centrality and system, as mentioned previously. In the right panel we observe that the strangeness saturation factor ($\gamma_{S}$), which characterises the deviation in strangeness production from thermal expectations, appears to lie on a common trend with respect to $\langle N_{part} \rangle$. Strangeness saturation i.e. $\gamma_{S}=1$ in both Cu+Cu and Au+Au appears to be reached at $\langle N_{part} \rangle \sim100$ in this scheme.
It is interesting to note that despite the extra strangeness production in central Cu+Cu collisions at given $\langle N_{part} \rangle$, $\gamma_{S}$ appears to be the same. 
Since the mid-rapidity $K/\pi$ ratio has been shown to be consistent for Cu+Cu and Au+Au at a given \Npart \cite{AnetaPri}, this indicates that $\gamma_{S}$ is controlled by the relative rates of non-strangeness to strangeness production. It was also shown that the larger strange and non-strange yields in central Cu+Cu at a given $\langle N_{part} \rangle$ are accompanied by larger freeze-out volumes in the THERMUS model \cite{JunsPres}.

\section{Mid-$p_{T}$ Spectra}

\begin{figure}[h]
\begin{tabular}{c c}
\includegraphics[width = 0.47\textwidth]{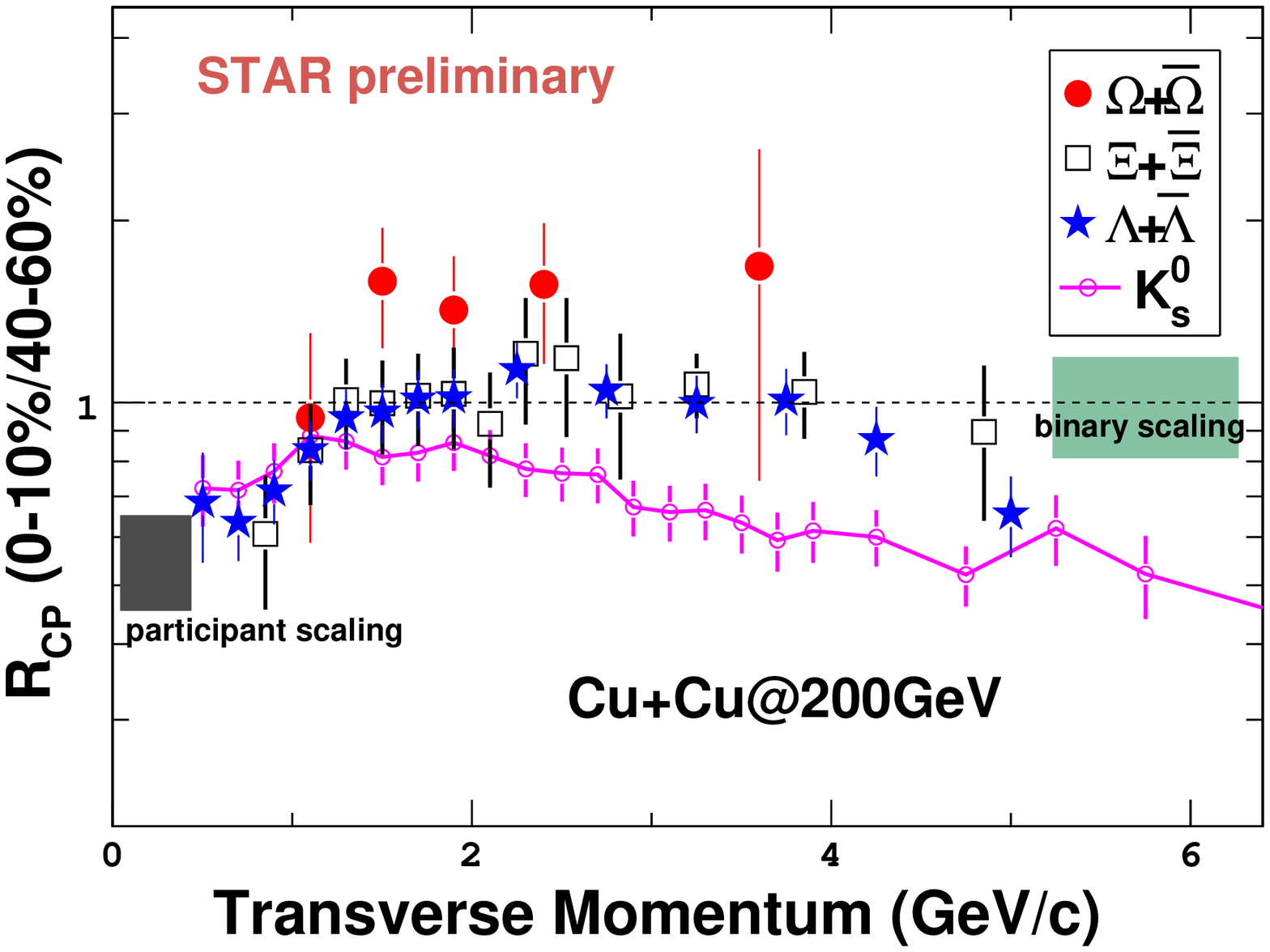}
&
\includegraphics[width = 0.51\textwidth]{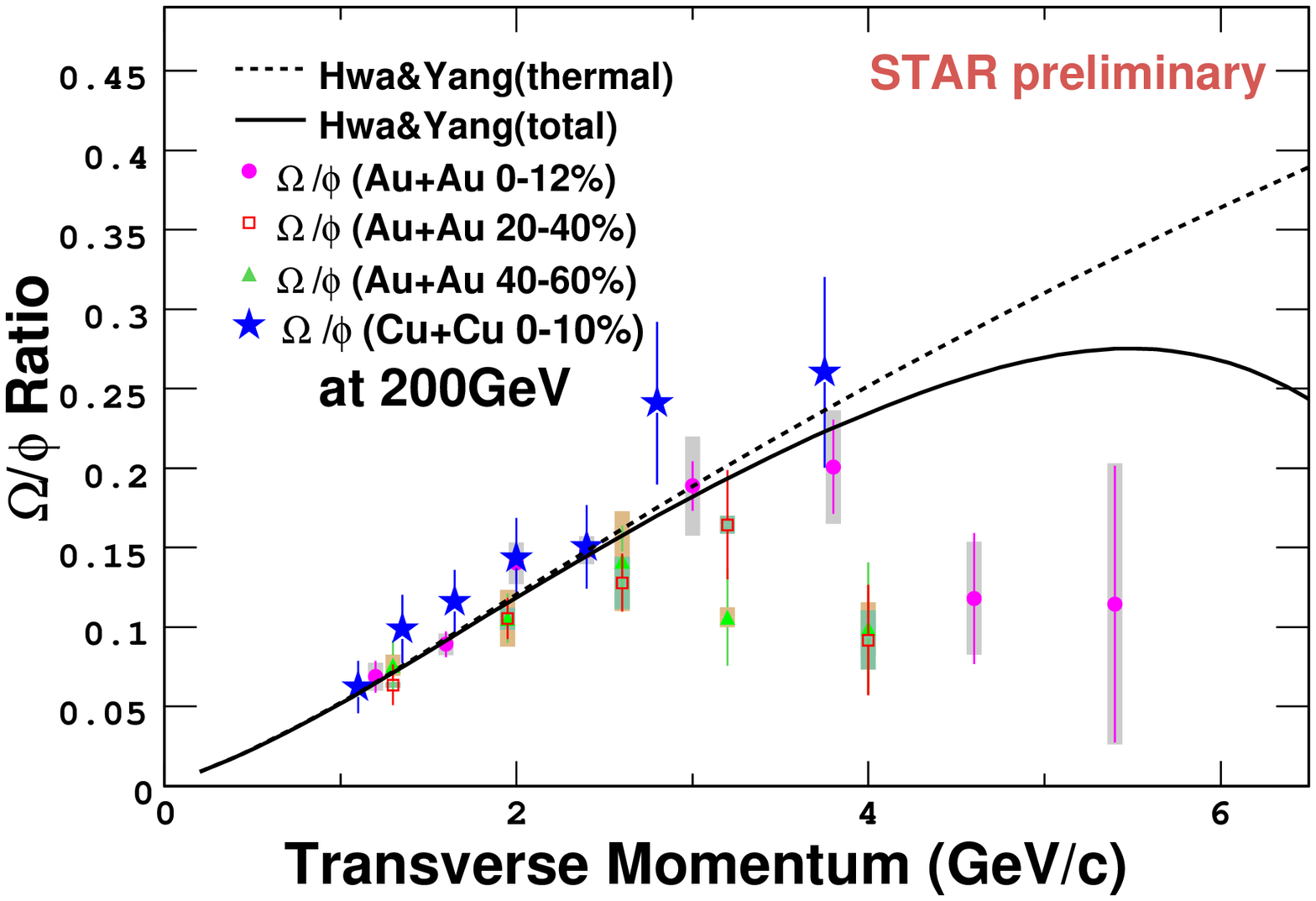}
\end{tabular}
\caption{Left Panel: $R_{CP}$ for strange mesons and baryons in Cu+Cu \sNN{200} collisions. The shaded boxes show regions where the respective scaling regimes apply. Right Panel: The $\Omega/ \phi$ ratio in Cu+Cu and Au+Au \sNN{200} collisions. The uncertainties are statistical in both panels \cite{XWang, PhiOmega}.}
\label{fig:4}
\end{figure}

The left panel of figure \ref{fig:4} shows $R_{CP}$ for various strange particles in Cu+Cu \sNN{200} collisions. This is defined as the yield per binary collision in central collisions divided by respective value for peripheral collisions. The mid-$p_T$ ($2 <  p_{T} <  4$ GeV/c) baryons have higher values compared to the strange meson in this region, and this is often linked to coalescence providing an extra source of particle production for baryons relative to mesons. This was also observed in Au+Au \sNN{200} collisions \cite{MattSQM}. The values at higher $p_T$ appear to merge where jet fragmentation is expected to dominate particle production. The $R_{CP}$ values below one indicate jet energy loss in central Cu+Cu collisions and they are consistent with the charged hadron values in the same region \cite{PHOBOSRAA}. The right panel in figure \ref{fig:4} shows the  $\Omega/ \phi$ ratio for various centralities in  Cu+Cu and Au+Au \sNN{200} collisions. The dotted lines shows predictions for production from just coalescence, while the solid line shows predictions for coalescence and jet fragmentation. The inclusion of fragmentation leads to the observed turn over in Au+Au albeit at a different $p_T$. The favourable comparisons with theory at mid-$p_T$ again point to coalescence being the likely mechanism for $\Omega$ and $\phi$ production in this $p_T$ range for both Cu+Cu and Au+Au \sNN{200} collisions. Furthermore, $v_{2}$ for the $\Lambda$ and $K^{0}_{S}$ particles in both systems has been observed to follow constituent quark scaling, which is also indicative of coalescence \cite{HiroshiMas, v2AuAu}.

\section{$v_{2}$ Measurements}

$v_{2}$ measurements characterise the conversion from the initial state anisotropy  to momentum anisotropy, and thus may give information on the degree of rescattering in heavy-ion collisions. By implementing the Boltzman equation, transport theory can determine the degree of non-equilibrium behaviour \cite{Transport}. This is quantified via the Knudsen number $K$ which is the ratio of the mean free path length to the system size. Non-equilibrium $v_{2}$ can be expressed as follows:
\begin{equation}
\label{equ:Knud}
\frac{v_{2}}{\epsilon}  =  \frac{v_{2}^{hydro}}{\epsilon} \frac{1}{1+K/K_{0}}
\end{equation}
where $\epsilon$ is the initial state eccentricity and $K_{0}$ a constant. When $K=0$, $v_2$ is fully hydrodynamic like, and when $K>0$, $v_2$ is below hydro expectations as the effects of a finite mean free path and/or finite system size are felt by the system.
\begin{figure}[h]
\begin{center}
\includegraphics[width = 0.7\textwidth]{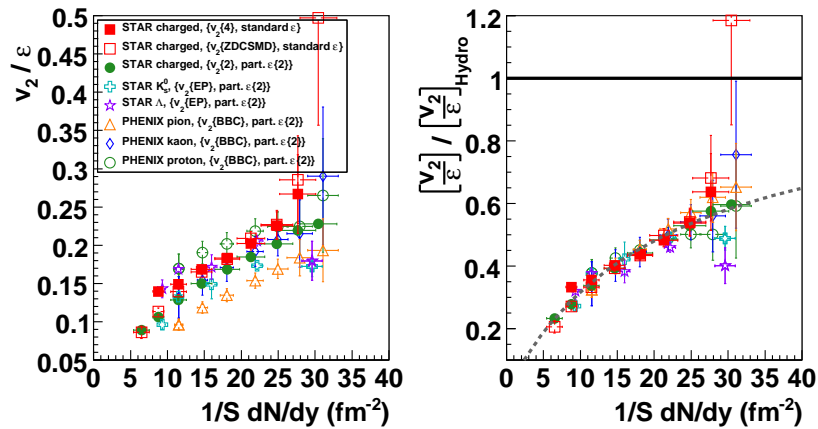}
\end{center}
\caption{Left Panel: Various integrated $v_2$ divided by eccentricity (y-axis) as a function of hadronic areal density (x-axis). Right Panel: Deviation in integrated $v_2$ (y-axis) from hydro expectations as a function of hadronic areal density (x-axis). A deviation of one indicates the measured $v_2$ corresponds to hydro expectations \cite{HiroshiMas, YuBai}.}
\label{fig:5} 
\end{figure}
The left panel of figure \ref{fig:5} shows $p_T$ integrated $v_2$ for strange and non-strange in Cu+Cu and Au+Au \sNN{200} collisions. The right panel shows the same data divided by the expected hydro value at the respective system density. The fitted transport curve shows the Knudsen number decreases with increasing density and does not reach zero for the highest densities where the measured $v_{2}$ is $\sim30\%$ below hydro expectations. Amongst other things, the apparent contradiction with earlier claims of ideal hydro behaviour results from eccentricity being better understood in heavy-ion collisions \cite{RaimondSnell}.
%
%\begin{figure}[h]
%\begin{center}
%\includegraphics[width = 0.5\textwidth]{Figure6}
%\end{center}
%\caption{Strange particle $v_2$ divided by the number of constituent quarks as a function of transverse kinetic energy per consituent quarks for various centralities in Cu
%+Cu \sNN{200} collisions.}
%\label{fig:5} 
%\end{figure}

\section{Low Energy Scan}
Various lattice QCD calculations predict a critical point where the QGP may hadronise via a first order phase transition \cite{Kumar}. This motivates a possible low energy scan ($ \sqrt{s_{NN}}\sim5 \rightarrow 40$ GeV) at RHIC that aims to cover the most recent predictions of temperature and baryo-chemical potential where the critical point may occur. Furthermore, as a collider experiment, STAR's constant acceptance with respect to beam energy gives it an advantage over its fixed target 
\begin{figure}[h]
\begin{center}
\includegraphics[width = 0.5\textwidth]{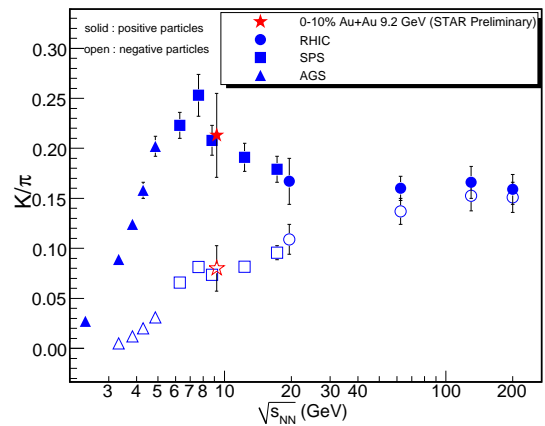}
\end{center}
\caption{Mid-rapidity $K^{+} / \pi^{+}$ and $K^{-} / \pi^{-}$ ratios for central Au+Au \sNN{9.2} collisions. The uncertainties are statistical and systematic added in quadrature \cite{Kumar}.}
\label{fig:6}
\end{figure}
counterparts where other searches maybe carried out. Earlier this year, STAR recorded $\sim3000$ events from Au+Au \sNN{9.2} collisions as a feasibility test for low energy running. The results are shown in figure \ref{fig:6} for the $K^{+}/\pi^{+}$ and $K^{-} / \pi^{-}$ ratios.
It is clear that STAR has already achieved global consistency with other experimental results. The same is found to be true for the integrated mid-rapidity $K^{-}/ K^{+}$, $\pi^{-}/ \pi^{+}$, and $p^{-}/ p^{+}$ ratios, and the $K^{+} / \pi^{+}$ and $K^{-} / \pi^{-}$ ratios from Au+Au \sNN{19} collisions recorded in 2000 \cite{Kumar}.

\section{Strange Particle Fragmentation Functions in p+p}

\begin{figure}[h]
\begin{center}
\includegraphics[width = 1\textwidth]{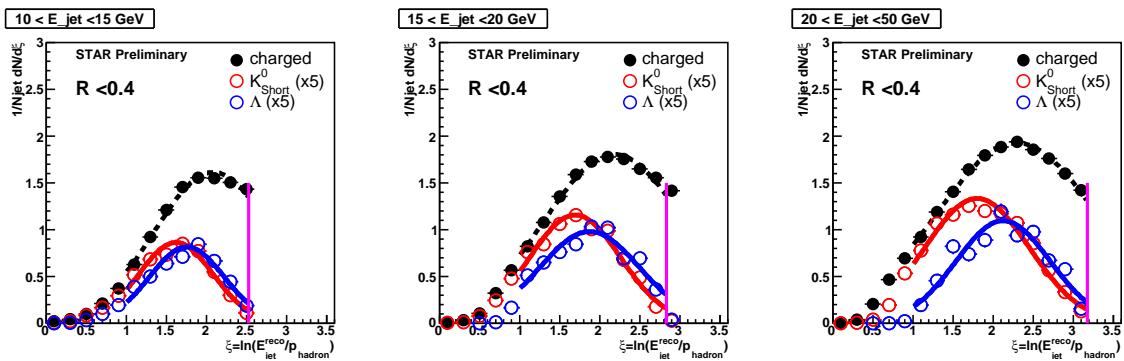}
\end{center}
\caption{Fragmentation functions of charged hadrons, and the \Lam and \Kaon particles for various jet energies. $R$ indicates the size of jet the cone in $\Delta \eta, \Delta \phi $ space. The cut off at high $\xi$ is due to limitations in the jet finding algorithm. The lines are Gaussian fits \cite{MarkHeinz}.}
\label{fig:7} 
\end{figure}

We finally turn to strange particle fragmentation functions measured in p+p \sNN{200} collisions. These provide an essential reference for understanding strangeness production from jets in heavy-ion collisions, and offer constraints to various QCD inspired jet fragmentation models. Figure \ref{fig:7} shows fragmentation functions for the charged hadron, $\Lambda$, and \Kaon particles at three different jet energies. The fragmentation function is expressed in terms of $\xi=ln(E_{jet}/p_{T}^{hadron})$ where $E_{jet}$ is the reconstructed jet energy, and the respective probability function is normalised by the number of jets in the energy range studied. For a particular particle species, the integral of the function gives the mean number per jet over a chosen range in $\xi$.
There are two key observations to be made from the data in figure \ref{fig:7} for all jet energies. Firstly, it was found that for the particle species studied, the mean $\xi$ violated a predicted mass ordering expected in QCD \cite{MassOrder}. Secondly, for a hadron of $p_T \sim 2$ GeV/c, it was found that the $\Lambda /K^{0}_{S}$ ratio $\sim1$. Since p+p events with higher values of $N_{charge}$ are more likely to contain a hard scattered parton \cite{ppStudies}, this observation is supported by the fact that the inclusive mid-$p_T $ $\Lambda /K^{0}_{S}$ ratio approaches one with increasing $N_{charge}$ \cite{STARpp}. Taken at face value, this readdresses the question of exactly what is a jet-like baryon/meson ratio, with the answer being crucial for heavy-ion studies in the mid-$p_{T}$ range. Finally, it must be noted that the accuracy of each of the measurements behind the respective observations is still under evaluation.

%\begin{figure}[h]
%\begin{center}
%\includegraphics[width = 0.5\textwidth]{Figure8}
%\end{center}
%\caption{ds.}
%\label{fig:8} 
%\end{figure}

\section{Summary}

We have presented recent results for strange particle $dN/dy$ in Cu+Cu and Au+Au collisions and found for the collision energy of \sNN{200}, both the EPOS and AMPT models reproduce qualitative features of the data with EPOS doing better quantitatively. It was shown that baryon-meson differences occur for $R_{CP}$ in Cu+Cu \sNN{200} collisions, and that a recombination model describes the $p_T$ dependance of the $\Omega/\phi$ ratio for central Cu+Cu collisions at the same energy. Strange and non-strange particle $v_2$ were found to exhibit non-equilibrium behaviour for Cu+Cu and Au+Au \sNN{200} at all centralities, and STAR's recent low energy run gave $K/\pi$ ratios consistent with other experiments. Strange particle fragmentation functions were presented for p+p \sNN{200} collisions, and these showed a breaking of the expected mean $\xi$ mass ordering and a mid-$p_{T}$ ($\sim 2$ GeV/c) $\Lambda /K^{0}_{S}$ ratio of $\sim1$ in jets for the kinematic range studied.

\end{document}